\documentclass{Indulekha} 
\usepackage{graphicx}

\title[JD 11.~~Density dependent IMF and progenitors of GW sources] 
{Implications of a density dependent IMF \\ for the statistics of  \\ progenitors of gravitational wave sources}

\author[Indulekha Kavila \& Megha Viswambharan]   
{Indulekha Kavila
 \and Megha Viswambharan}

\affiliation{School of Pure \& Applied Physics, Mahatma Gandhi University, Kottayam 686560 INDIA \\ email: {\tt indulekha@mgu.ac.in}\\email: {\tt meghapv7@gmail.com}}

\pubyear{2018}
\volume{346}  
\jname{High-mass X-ray binaries: illuminating the passage from
massive binaries to merging compact objects}
\editors{L. Oskinova, D. Gies, E. Bozzo \& T. Bulik, eds.}

\begin{document}

\maketitle

\begin{abstract}
Observations of mergers of multi-compact object systems offer insights to the formation processes of massive stars in globular clusters.  Simulations of stellar clusters, may be used to understand and interpret observations.  Simulations generally adopt an Initial Mass Function (IMF) with a Salpeter slope at the high mass end, for the initial distribution of stellar masses.  However, observations of the nearest high mass star forming regions point to the IMF at the high mass end being flatter than Salpeter, in regions where the stellar densities are high.  We explore the impact of this on the formation rate of potential GW sources, estimated from standard considerations. Globular clusters being significant contributors to the ionization history of the universe, the results have implications for the same.  It impacts our ability to explore the putative mass gap, between the upper limit for neutron star masses and the lower limit for black hole masses, also.  
\keywords{binaries: close, globular clusters: general, open clusters and associations: general, early universe}
\end{abstract}

\firstsection 
\section{Introduction}

Gravitational wave signals are clean signals.  With the detection of gravitational wave (GW) signals from coalescing binary compact objects by LIGO (\cite [Abbott et al. (2016a, 2016c, 2017a, 2017b, 2017c, 2017d)]{Abbott_etal2016a}), precise determination of the masses and effective spin parameters of the coalescing objects have now become possible (\cite [Veitch et al. (2015)]{Veitch2015}; \cite [Abbott et al. (2016b)]{Abbott_etal2016b}).  The masses obtained so far are of stellar order.  These observations are thus probes for the nature and extent of the earliest processes of formation of Pop II stars. The length of the time interval expected, between binary formation and ultimate coalescence of the compact objects, suggests that, the progenitors of these GW sources formed very early. The oldest stars in galaxies are found in the globular clusters.  Globular clusters are putative hosts for GW sources, due to the possibility of formation of binary compact objects by capture (\cite [Abbott et al. (2016d)]{Abbott_etal2016d}).  Massive stars which formed in globular clusters are considered to have contributed to the reionization of the universe also (\cite [Boylan-Kolchin (2018)]{Boylan-Kolchin2018}).

	The effective spin parameters of binary compact objects, which formed through dynamical interactions of evolved objects in dense stellar environments like those in globular clusters, are expected to be zero, since there will no preference for alignment between the directions of orbital and spin angular momenta (\cite [Abbott et al. (2016d)]{Abbott_etal2016d}).  With a bunch of six detections of GW signals in hand, the results for all the sources are consistent with their effective spin parameters being zero.  However, the results are found to be not insensitive to the prior.  It has been noticed that Bayesian priors have an impact on the characterization of Binary Black Hole (BBH) mergers and that in the case of GW 151226, the odds are almost similar for a prior peaked around alignment for spin direction as for one isotropic in spin direction (\cite [Vitale et al. (2017)]{Vitale2017}).    An alternative to the formation of binary compact objects by capture, is the scenario where a binary evolves through common envelope evolution, surviving the two supernova explosions (\cite [van den Heuvel E P J (1981)]{vandenHeuvel1981}).  In this context, recent observations of non-standard results regarding the IMF in high mass star forming regions and intra-core velocity dispersion in dense, star forming cores, are analyzed to understand their impact on the formation rate of massive binaries. 

\section{Analysis and results}

\begin{figure}[b]
\begin{center}
 \includegraphics[width=3.4in]{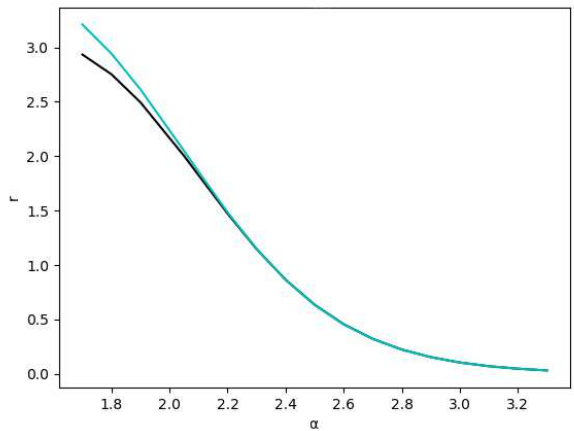} 
 \caption{The ratio $r$ of the number of stars with mass $>$ 8 solar masses expected for sampling from an IMF with slope $\alpha$ with the number expected from sampling from an IMF with Salpeter slope for a cluster of given mass.  The dark line is for the case where the upper limit for stellar masses is taken as 150 solar masses and the lighter one is for an upper mass limit of 100 solar masses.}
   \label{fig1}
\end{center}
\end{figure}

A high star density and a preponderance of massive stars have been both hypothesized as well as observed in the case of massive star clusters and starbursts, where star formation takes place in assemblies of dense gas.  Examples include the Milky Way clusters Arches, Westerlund 1 and NGC 3603 (see for eg \cite[Habibi et al. (2013)]{Habibi_etal2013}; \cite [Lim, Sung \& Hur (2014)]{LimSungHur2014};  \cite [Harayama, Eisenhover \& Martins (2008)]{Harayama2008}).  The starburst region 30 Dor in the Large Magellanic Cloud is observed to have a flatter IMF and a preponderance of massive stars over and above the numbers expected from a Salpeter IMF (\cite [Schneider et al. (2018)]{Schneider2018}).  Sub-virial velocity dispersions have been observed for the denser component of gas in cores (\cite [Kirk et al. (2010)]{Kirketal2010}; \cite [Wilking et al. (2015)]{Wilking2015}).  Both the above factors can enhance the chances for the formation of  binary compact objects.  Only $\sim  10 \%$ of stellar clusters which form, survive bound, the rest dissolving into the field.  Smaller velocity dispersion within the denser gas of cores and locally nonstandard IMFs, both enhance the probability for the formation of field BBHs with aligned spins, over and above the expectation from the standard scenario. This enhancement factor is estimated.  Significant spin alignment of stars has been noticed in observations of old open clusters (\cite[Corsaro et al. (2017)]{Corsaro_etal17}).  Stellar clusters do show a hierarchical sub-clustering which is interpreted as the imprint of the density distribution of the supersonic turbulence in the gas from which they formed via fragmentation (\cite[Gouliermis et al. (2014)]{Gouliermis2014}).  Dense substructures thus offer  conditions that are favorable for the formation of binaries.

Enhancement of the probability of formation of massive binaries due to non-standard IMF:  For a given mass of stars, the ratio $r$ of the number of stars with mass greater than $8$ solar mass, sampled from an IMF with slope $-\alpha$, with that sampled from an IMF with Salpeter slope, is determined (Fig.\,\ref{fig1}).  The ratio $r$ is $\sim  2$ for flatter IMFs with $\alpha  > 2$.  The Salpeter value for the slope is $-2.35$.

Enhancement of the probability of formation of massive binaries due to sub-virial intra-core velocity dispersion:  If the expected relative speeds of protostars are only, say, a factor $a$ times the expected virial speed, the probability for binary formation by capture, during the early phases of star formation will be higher by a factor $a^{-3}$.  Dispersion in the intra-core line-of-sight (los) velocities of dense gas is found to be closer to thermal than virial in dense cores.  The observed mean dispersion in the los velocity is $< 0.5 $\,\rm{km/s} where the velocity dispersion of the surrounding less dense gas is of the order of $1$ - $2  $\,\rm{km/s} (\cite [Kirk et al. (2010)]{Kirketal2010}; \cite [Wilking et al. (2015)]{Wilking2015}).

It may be seen from the above that, taking into account the possible non-standard nature of the IMF in dense high mass star forming regions and the low intra-core velocity dispersions, enhances the chances for the formation of BBHs by an order of magnitude.  

\section{Discussion}
A flatter IMF for star formation taking place at high redshifts has been proposed to account for the inconsistency between the cosmic stellar mass density and the observed star formation rate up to z $\sim 8$ (\cite [Yu \& Wang (2016)]{YuWangng2016}).  The IMF of the nearby 30 Dor starburst region, which is considered to be an analogue of distant starbursts in the early universe, has a slope $\sim 1.9$, for stars having masses $>$ 15 solar masses (\cite [Schneider et al. (2018)]{Schneider2018}). It is pointed out that a flatter IMF, along with low intra star-forming-core velocities which could arise from the high dissipation rate of turbulence in the dense cores, may lead to at least an order of magnitude enhancement, of the chances for the formation of binary black holes with aligned spins, compared to the chances for the same in the standard scenario (\cite [Abbott et al. (2016d, 2016e)]{Abbott_etal2016d}; \cite [Kalogera (2000)]{Kalogera2000}).  

Globular clusters are the oldest objects accessible for observation in the nearby universe. The advent of next generation telescopes like the James Webb Space Telescope and the Thirty Meter Telescope, will allow detailed exploration of binaries, the binary fraction, and the evolution of both, in crowded and / or deeply embedded star burst fields as well as globular cluster fields.  With the advent of aLIGO (Advanced LIGO) and IndIGO (Indian Initiative in Gravitational-wave Observations) it will be possible to locate GW sources in the sky, enabling multi-messenger astronomy of binary black hole mergers and identification of the hosts of the BBHs.  Simulations allow us to backtrack from observations to the formation and evolution of binaries and the binary fraction of stellar clusters.  

The formation of massive stars is still a mystery.  The above considerations show that observations of multi-compact object mergers offer insights to massive star formation in the early universe.  Considering the fact that massive stars are significant contributors to the ionization history of the early universe, the study of the details of the formation and subsequent events in the lives of massive multi-star systems attains great importance.

\end{document}